\documentclass{emulateapj}

\usepackage{mathrsfs}

\newcommand{\etal}{{et al.}}            

\newcommand{\XMM}{{\em XMM-Newton }}
\newcommand{\Ch}{{\em Chandra }}
\newcommand{\ot}{[O~{\sc iii}] }

\def\gappeq{\mathrel{ \rlap{\raise.5ex\hbox{$>$}}
                      {\lower.5ex\hbox{$\sim$}}  } }
\def\lappeq{\mathrel{ \rlap{\raise.5ex\hbox{$<$}}
                      {\lower.5ex\hbox{$\sim$}}  } }

\shorttitle{X-RAY SPECTROSCOPY OF THE NGC 2110 NUCLEUS}
\shortauthors{EVANS ET AL.}

\begin{document}

\title{Probing Unification With Chandra HETGS and XMM-Newton EPIC And RGS Spectroscopy of the Narrow Emission Line Galaxy NGC 2110}
\author{Daniel~A.~Evans\altaffilmark{1}, Julia~C.~Lee\altaffilmark{1}, T.~Jane~Turner\altaffilmark{2,3}, Kimberly~A.~Weaver\altaffilmark{2}, and Herman~L.~Marshall\altaffilmark{4}}
\altaffiltext{1}{Harvard-Smithsonian Center for Astrophysics, 60 Garden Street, Cambridge, MA 02138}
\altaffiltext{2}{NASA Goddard Space Flight Center, Code 662, Greenbelt, MD 20771}
\altaffiltext{3}{Dept. of Physics, University of Maryland Baltimore County, 1000 Hilltop Circle, Baltimore, MD 21250}
\altaffiltext{4}{MIT Kavli Institute for Astrophysics and Space Research, 77 Massachusetts Avenue, NE 80, Cambridge, MA 02139}

\begin{abstract}

We present results from {\it Chandra} HETGS (250 ks over two epochs) and {\it XMM-Newton} EPIC and RGS (60 ks) observations of NGC~2110, which has been historically classified as a Narrow Emission Line Galaxy galaxy. Our results support the interpretation that the source is a Seyfert~2 viewed through a patchy absorber. The nuclear X-ray spectrum of the source is best described by a power law of photon index $\Gamma\sim1.7$, modified by absorption from multiple layers of neutral material at a large distance from the central supermassive black hole. We report the strong detections of Fe~K$\alpha$ and Si~K$\alpha$ lines, which are marginally resolved with the \Ch HETGS, and we constrain the emission radius of the fluorescing material to $\gappeq1$ pc. There is some evidence for modest additional broadening at the base of the narrow Fe~K$\alpha$ core with a velocity $\sim4500$~km~s$^{-1}$. We find tentative evidence for ionized emission (O~{\sc viii}~Ly~$\alpha$, an O~{\sc viii}~RRC feature, and possibly a Ne~{\sc ix} forbidden line) in the \Ch MEG and \XMM RGS spectra, which could be associated with the known extended X-ray emission that lies $\sim$160 pc from the nucleus. We suggest that the $10^{23}$~cm$^{-2}$ partially covering absorber originates in broad-line region clouds in the vicinity of the AGN, and that the $3\times10^{22}$~cm$^{-2}$ coverer is likely to have a more distant origin and have a flattened geometry in order to allow the small-scale radio jet to escape.

\end{abstract}


\section{INTRODUCTION}

The discovery of broad emission lines in the polarized spectra of a number of Seyfert 2 galaxies (\citealt{ant85}) has laid the framework for Unified models of Active Galactic Nuclei (AGN) (e.g.,~\citealt{ant93,up95}). In the simplest Unification schemes, broad optical line objects (typically classed as Seyfert 1 galaxies) and narrow optical line objects (Seyfert 2 galaxies) are intrinsically identical, and the main discriminator between them is the angle to the line of sight made by an optically thick `torus' that surrounds the nucleus. While these models have had remarkable successes in explaining the observed properties of Seyfert 1 and Seyfert 2 galaxies, there remains some contrary evidence (e.g., \citealt{mal98,tran01,buc06}) that cautions skepticism about Unified models, at least in their simplest forms. One interesting example is the study of Narrow Emission Line Galaxies (NELGs), the subclass of Seyfert galaxies that were first identified at X-ray energies due to their strong and flat-spectrum ($\Gamma$$\sim$1.5) 2--10 keV X-ray emission (e.g.,~\citealt{mus82,rom96}). Indeed, NELGs were proposed to contribute significantly to the flat, low energy portion of the X-ray background (e.g.,~\citealt{gen95}). Although the optical spectra of NELGs is dominated by narrow ($<600$km s$^{-1}$) emission lines (\citealt{shu80}), consistent with Seyfert 2 galaxies, a number of them have broad H$\alpha$ emission, which is more reminiscent of Seyfert 1 sources. This led \cite{law82} to suggest that NELGs may represent transition objects between the two classes.

X-ray spectroscopy of the continuum, fluorescent emission lines, and absorption features in the nuclei of Seyfert galaxies is an important probe of the geometry of AGN. In particular, studies of the Fe~K$\alpha$ line in Seyfert galaxies with the \Ch High Energy Transmission Grating Spectrometer (HETGS) have shown that its profile tends to be complex, with a narrow ``core'' sometimes accompanied by broader emission (e.g.,~\citealt{yaq04}). If the contributions to the Fe~K$\alpha$ line complex from the narrow and broad components can be accurately deconvolved, important constraints can be placed on the inclination of the AGN with respect to the observer, thereby providing a test of Unified schemes. While the narrow core is attributed to fluorescent emission at a large distance ($\gappeq$1,000 $R_{\rm g}$) from the central black hole, plausibly in the form of a torus, the origin of any broad component that might be present remains ambiguous. For example, it could be a signature of the relativistic inner portion of the accretion flow, have an origin in the broad line region (BLR), or simply be an artifact of inadequately modeling the circumnuclear absorption or not adopting a self-consistent model of reflection in terms of line and continuum features.

NGC~2110, the subject of this paper, is a nearby flat spectrum ($\Gamma$$\sim$1.5) NELG a redshift $z$=0.00801 ($D_{\rm L}$=34.5~Mpc). {\it HEAO-1} observations of the source ({\citealt{mus82}) found an absorbed ($N_{\rm H}\sim7\times10^{22}$ cm$^{-2}$) power-law spectrum, accompanied by short-timescale variability in hard X-rays, suggesting that the hard X-ray emission originates within 0.2 pc of the central engine. \cite{hay96} demonstrated with {\it Ginga} and {\it ASCA} observations of NGC 2110 that the 2--10 keV X-ray spectrum is relatively flat, is accompanied by an Fe~K$\alpha$ line, and that reflection is not dominant in this object. {\it BeppoSAX} observations (\citealt{mal99}) confirmed the 2--10 keV spectrum is rather flat ($\Gamma$$\sim$1.5--1.6), but showed that it steepens to $\Gamma=1.9$ when data above 13 keV are considered, consistent with typical values ($\Gamma$=1.7--2.0) found in Seyfert type 1 sources (e.g.,~\citealt{nan94}). In short, there is evidence that both Seyfert~1 (hard X-ray luminosity, fast time variability) and Seyfert~2 (broad H$\alpha$ line emission, large absorption at low energies) characteristics are present in the nuclear spectrum of NGC~2110.

Further insights into the Seyfert classification of NGC~2110 come from observations of its Fe~K$\alpha$ line. Based on {\it ASCA} observations of a sample of type 2 Seyfert galaxies (including NGC~2110), \cite{tur98} argued that the observed Fe~K$\alpha$ line profile is broadened and so consistent with an accretion-disk origin, although narrow enough that the disk is oriented nearly face-on with respect to the observer, with an inclination comparable to type 1 Seyfert sources. An alternative model (\citealt{wea98}) interprets the line profile as the sum of an intrinsically narrow component, attributed to the circumnuclear absorber, and a broadened ``diskline'' inclined at an angle of 48$^{\circ}$ to the observer. This model is consistent with Unified AGN schemes in which Seyfert 2 galaxies are inclined at low (i.e., edge-on) to intermediate angles with respect to the observer. However, the CCD-quality spectral resolution of {\it ASCA} was likely insufficient to allow the deconvolution of broad/narrow Fe~K$\alpha$ line emission and complex absorption effects, and so we aim to use the superior energy resolution of the \Ch HETGS instrument in particular to provide new insights to the geometry of the nucleus of NGC~2110.

In addition to the debate surrounding the Fe~K$\alpha$ line, there remain several interpretations for the origin of the soft X-ray excess, first seen in the spectrum of NGC 2110 at energies below 1 keV (e.g.,~\citealt{mus82}). \cite{evans06} presented a multiwavelength analysis of NGC~2110, using data from {\it Chandra}, {\it HST}, and VLA imaging observations of the circumnuclear extended emission in the source. We found resolved soft-band X-ray emission $\sim4''$ ($\sim160$ pc) north of the nucleus, which is spatially coincident with resolved \ot and H$\alpha$ emission, but lies just beyond the northern edge of the radio jet. We considered the energetics of several different physical scenarios that may account for the extended X-ray emission, and demonstrated that it is possible that the radio jet has driven a strong shock through a series of multi-phase clouds of gas. Alternative models, such as the scattering of nuclear radiation by a population of electrons, or the photoionizaton of ambient gas by the nucleus, may also be responsible for the extended emission. The high spectral resolution of the \Ch HETGS and \XMM RGS instruments enable us to search for characteristic signatures of collisionally ionized and photoionized plasmas, and place important constraints on the nature of the soft X-ray emission. Indeed, as part of an archival \XMM RGS study of 69 obscured AGN,  \cite{gua06} detected a narrow ($\sigma<3.2$~eV) O~{\sc viii} Radiative Recombination Continuua (RRC) feature in NGC~2110, suggesting that photoionization may be important in the source.

In this paper, we present the results of \Ch HETGS and \XMM EPIC and RGS observations of the continuum, reflection spectrum, absorption features and ionized lines in the nuclear spectrum of NGC~2110, and use this to constrain the geometry of the source and therefore test AGN unification models. In Section 2, we describe the data and summarize their reduction. In Section 3, we present a variability analysis of NGC~2110. In Section 4, we provide an initial parametrization of the broad-band continuum of the source, while in Section 5, we present high-resolution \Ch HETGS spectroscopy of the fluorescent reflection spectrum. We present evidence for excess photoelectric absorption at Fe and Si in Section 6, which motivates us to examine a more complex model for the X-ray continuum in Section 7. We discuss the low-energy ($<$1~keV) emission spectrum in Section 8, and briefly show the CCD spectrum of a contaminating point source in Section 9. In Section 10, we interpret the results in terms of the geometry of the AGN and the origin of the reflection spectrum. We end with our conclusions in Section 11.

All results presented in this paper use a cosmology in which $\Omega_{\rm m, 0}$ = 0.3, $\Omega_{\rm \Lambda, 0}$ = 0.7, and H$_0$ = 70 km s$^{-1}$ Mpc$^{-1}$. All spectral fits include absorption through our Galaxy using $N_{\rm H, Gal}$=$1.76\times10^{21}$ cm$^{-2}$ (\citealt{dic90}), although an alternative value of $1.82\times10^{21}$~cm$^{-2}$ was measured by \cite{elv89}. The energies of any emission and absorption features are quoted in the source rest frame. Errors quoted in this paper are 90 per cent confidence for one parameter of interest (i.e., $\chi^2_{\rm min}$ + 2.7), unless otherwise stated. When distinguishing between different model fits to the data, we consider $F$-statistic results, and we adopt a threshold of 95\%  for a significant improvement in the fitting statistic.

\section{OBSERVATIONS AND DATA REDUCTION}

NGC~2110 was observed with the \Ch High Energy Transmission Grating Spectrometer on four occasions: the first three observations (PI J.~C.~Lee) were performed within three days of each other in 2001 December (OBSIDs 3143, 3417, and 3418), while the fourth (P.I. T.~J.~Turner) was performed on 2003 March 05 (OBSID 4377). NGC~2110 was observed with \XMM EPIC and RGS instruments on 2003 March 05 (OBSID 0145670101, PI T.~J.~Turner), as part of a simultaneous observation of the source with {\it Chandra} (OBSID 4377). A summary of the observations is given in Table~\ref{obslog}. In Sections~\ref{hetgsreduction},~\ref{epicreduction}, and~\ref{rgsreduction} we describe how we reduced the HETGS, EPIC, and RGS data, respectively.

\subsection{Chandra HETGS}
\label{hetgsreduction}

The 2001 December set of \Ch HETGS observations were performed in an interleaved mode, with 3 frames of 3.2 seconds alternated with 1 frame of 0.8 seconds. However, alternate-frame data processed with versions of the Standard Data Processing software prior to v. DS 6.9 contain an error in the pipeline derivation of the DTCOR (=LIVETIME/ONTIME) keyword, resulting in incorrect exposure times. We therefore requested a special reprocessing of the data by the \Ch Science Center using the latest SDP software in order to resolve this error. The 2003 March observation was performed with a 2.5~s frame exposure time, and no special reprocessing was necessary.

Before extracting the gratings spectra, we inspected images of the dispersed X-ray spectra in order to search for unrelated X-ray sources. We found that the MEG+1 dispersion axis for the \Ch observation performed in 2003 March (OBSID 4377) is contaminated by an unrelated soft X-ray source on the ACIS-S4 chip at $\sim9''$ from the nucleus, corresponding to a dispersion energy of $\sim$1~keV. This source was masked from our analysis. We report on this source in Section~\ref{unrelatedps}.

Spectra and instrument responses were generated using {\sc CIAO} {\sc v}3.3.0.1 and analyzed using the {\sc ISIS} spectral fitting software (\citealt{hou02}). For improved accuracy, we used an internal-release version of the OSIP file, which is now publicly available in the latest version of {\sc CIAO}, {\sc v}3.4. The $+1$ and $-1$ orders of the HEG were combined, as were the $+1$ and $-1$ orders of the MEG. For the 2003 December series of exposures, this method was applied individually to the 0.8-second frame time data and 3.2-second frame time data, resulting in 6 sets of spectra and response files for the HEG and MEG, before co-adding the 0.8 s and 3.2 s data for each observation to leave three datasets for each of the HEG and MEG. The 2003 March HETGS observations were not performed in interleaved mode.

Initially the four datasets were analyzed separately to search for temporal variability in either line or continuum features. Our analysis showed the spectra of the three 2001 December observations to be consistent with each other. In Section~\ref{hetgs_variability}, we demonstrate that the 2003 March data were consistent with the 2001 December data in terms of spectral shape, and varied {\it only} in terms of overall luminosity. This motivates our decision to combine the four HETGS datasets for our spectral analysis, which results in a total observation time of 209~ks for each of the HEG and MEG.

\subsection{XMM-Newton EPIC}
\label{epicreduction}

In our \XMM EPIC analysis we considered data from the pn camera only, which was operated in the Large Window mode with the thin optical blocking filter in place. The results presented here were obtained using the {\sc SAS} version 7.0.0. Calibrated event files were generated using the {\sc EPCHAIN} scripts.

To check for intervals of high particle background, light curves were extracted for the CCD on which the source is located, excluding large circle centered on the source itself. The events were filtered to include only those with PATTERN=0 and FLAG=0 attributes, and for an energy range of 10--12 keV. Inspection of the lightcurves revealed a significant impact from flaring. We adopted a conservative count-rate filtering criterion of 0.025 s$^{-1}$, which reduced the exposure time from 47.8 ks to 34.3 ks.

The \XMM observation of the nucleus of NGC~2110 is only marginally affected by photon pileup. We demonstrate this using two independent methods.  First, we determined the count rate in a 35$''$ source-centered circle with no PATTERN, FLAG, or energy filtering. The count rate is $\sim$3 s$^{-1}$, c.f. the maximum recommended count rate in this configuration to avoid pileup of 10 s$^{-1}$ (\citealt{xmmuhb}). Second, we used the SAS task {\sc epatplot} to determine the fractions of single- and double-pixel events as a function of energy, quantities which are sensitive to photon pileup. The task shows that the ratios are consistent with an observation free from pileup to energies $\lappeq$9 keV. Our analyses indicate that, although present, photon pileup is unlikely to significantly affect the results of our spectral analysis.

\subsection{XMM-Newton RGS}
\label{rgsreduction}

The \XMM Reflection Grating Spectrometer (RGS) data were reduced with SAS version 7.0.0, using the standard task {\sc rgsproc} to produce spectra and responses for the RGS1 and RGS2 cameras. Lightcurves of background events showed the RGS data to be affected by flaring in the same manner as the \XMM EPIC data, and we subsequently created a GTI table, which was used as an additional input to the task {\sc rgsproc} at the {\sc filter} entry stage, reducing the exposure time to 42 ks. In our analysis, data from the RGS1 and RGS2 cameras were {\it not} co-added.

\section{Variability}
\label{hetgs_variability}

In order to assess whether we could co-add the spectra from {\it all} four \Ch observations, we performed a simple variability analysis of the HETGS data. We extracted lightcurves from all four epochs of \Ch HETGS data using the {\sc aglc}\footnote{http://space.mit.edu/cxc/analysis/aglc/} script, for the soft (0.5--2 keV) and hard (2--7 keV) bands. The lightcurves, which comprise data from the +1 and -1 order of the HEG and MEG, are binned to 500-second time intervals and shown in Figure~\ref{hetgs_lightcurves}. The first three observations, performed in 2001 December, show no variability, and hence the data were combined for subsequent analysis. However, the fourth observation, performed in 2003 March, shows a significant decrease in count rate by the same factor in both the soft and hard bands.

Next, we searched for variability in the spectral shape of NGC~2110 by extracting spectra for the 2001 December \Ch HETGS observations (the three separate exposures were combined), and the 2003 March \Ch HETGS and \XMM EPIC observations. We fitted an absorbed power law to the three spectra in the energy range 2--9 keV, excluding the Fe~K$\alpha$ bandpass of 6.0--7.2 keV. Figure~\ref{contours_lightcurves} shows 90\% confidence (for two interesting parameters) contours for the intrinsic absorption ($N_{\rm H}$) and photon index ($\Gamma$) for the three sets of spectra. This illustrates that the continuum spectral parameters are consistent across the three datasets, and that the only observed variability is simply in the overall luminosity of the source. This motivates our choice to co-add the 2001 December and 2003 March HETGS data.

\section{Initial Parametrization of The Continuum And Soft Excess}
\label{initialspectralfitting}

We attempted an initial analysis of the continuum spectrum of NGC~2110 using the \Ch MEG spectrum (summed over all four epochs) and the \XMM pn spectrum. As we go on to describe in Sections~\ref{reflectionspectrum} and~\ref{absorptionfeatures}, the model fits described here do not represent the final parametrization of the continuum spectrum, necessitating a more complex model which we discuss in Section~\ref{finalmodel}. The data were grouped to a minimum of 100 counts per bin. We fitted an absorbed power law to the data over the energy range 2--9 keV, excluding the Fe~K$\alpha$ bandpass of 6.0--7.2 keV. We kept all parameters linked across both datasets, except for the normalizations of the power laws. The continuum was well fitted with a power law of photon index $\Gamma=1.45\pm0.05$ (but see Section~\ref{finalmodel}) and intrinsic neutral absorption of $N_{\rm H}=(2.89\pm0.16)\times10^{22}$ cm$^{-2}$. We then included data in the energy range 0.5--2 keV, which clearly shows a soft excess at energies $\lappeq1.5$ keV. Figure~\ref{softexcess_merged_MEG_pn_try1} shows the spectra and data/model ratio for both the MEG and pn data in the energy range 0.5--9 keV (note that the response of the MEG is poor above $\sim6$ keV), and shows the soft excess and Fe~K$\alpha$ line. It is clear from Figure~\ref{softexcess_merged_MEG_pn_try1} that the soft excess observed in the \XMM spectrum is greater than that in the \Ch spectrum. This is highly likely to be related to the presence of extended X-ray emission that our \Ch ACIS-S observation (\citealt{evans06}) reveals to lie $\sim30''$ south of the nucleus.

In order to model the soft excess, we added an unabsorbed power law, with its photon index linked to that of the heavily absorbed primary power law. Again, the Fe~K$\alpha$ bandpass of 6.0--7.2 keV was excluded from our continuum fits. The model provided a good fit to the data [$\chi^2=606.55$ for 698 degrees of freedom (dof)]. An improvement in the fit statistic ($\Delta\chi^2=9.08$ for one additional parameter) was achieved with the inclusion of mild neutral absorption [$N_{\rm H}=(4.16^{+2.37}_{-2.29})\times10^{20}$ cm$^{-2}$] on the soft power law, in excess of the Galactic column. In this case, the photon index is $\Gamma=1.40\pm0.04$ and the intrinsic absorption associated with the primary power law is $N_{\rm H}=(2.80^{+0.10}_{-0.09})\times10^{22}$ cm$^{-2}$. This improvement is significant at 99.9\% based on an $F$-test. Figure~\ref{softexcess_merged_MEG_pn_try3} shows the counts spectrum and data/model residuals for this fit, and Table~\ref{softexcess_merged_MEG_pn_try3_table} shows the best-fitting model parameters. Finally, we note that other models for the broad-band continuum, such as (1) the sum of a power law and a disk blackbody, both modified by the same cold absorber, or (2) a heavily absorbed power law plus a thermal ({\sc apec}) component, fail to fit the soft excess below $\lappeq$1 keV.

\section{The Reflection Spectrum}
\label{reflectionspectrum}

In our initial parametrization of the nuclear continuum, we find that the \Ch HETGS spectrum of NGC~2110 contains prominent fluorescent emission from neutral states of Fe and Si. We also report the additional detection of fluorescent S~K$\alpha$ emission, as well as tentative evidence for Ar and Ca K$\alpha$ lines. In Section~\ref{fe_si_narrowcores}, we describe in detail our analysis of the narrow Fe and Si~K$\alpha$ line cores, while in Section~\ref{evidenceforbroadening} we examine if there is accompanying broadened Fe~K$\alpha$ emission.

\subsection{The Fe and Si line cores}
\label{fe_si_narrowcores}

We used the \Ch HEG data to study features in the Fe~K bandpass. The data were grouped to a minimum of 4 channels per bin (bin size=0.01$\AA$) with at least 25 counts in each bin, in order to ensure the applicability of $\chi^2$ statistics. With these requirements, the data in the Fe~K bandpass is binned to the full instrumental resolution of the HEG. We fitted a heavily absorbed power law to the HEG data over the energy range 4--9 keV, ignoring for the time being the Fe~K bandpass of 6.0--7.2 keV. The continuum parameters were consistent at the 1$\sigma$ level with those found in Section~\ref{initialspectralfitting}. With the continuum parameters left to freely vary, we included the Fe~K bandpass in our spectral fitting and added a Gaussian of initial centroid energy 6.4 keV to represent the Fe~K$\alpha$ emission line, and allowed its energy, linewidth and normalization to vary freely. This resulted in a good fit to the data ($\chi^2=169.31$ for 154 dof, c.f., $\chi^2=184.06$ for 155 dof when the line width was frozen at an unresolved value). The energy of the line, $E=6.397\pm0.007$ keV (90\% confidence for two interesting parameters), is consistent with emission from neutral, or near-neutral species of iron. Figure~\ref{newfe_try2_contours} shows confidence contours of the energy and width of the Fe~K$\alpha$ line. The Fe~K$\alpha$ line is marginally resolved, with a width of $\sigma=19^{+13}_{-9}$~eV, c.f., the instrumental resolution of the HEG at this energy of $\sigma\simeq$14 eV. This corresponds to an intrinsic velocity width of $900^{+600}_{-400}$ km s$^{-1}$. The equivalent width of the line is $110\pm20$ eV, where the error quoted here takes into account only the 90\% confidence (for one interesting parameter) uncertainty on the normalization of the Gaussian. The addition of an Fe~K$\beta$ line width with free energy and normalization, and linewidth tied to that of the Fe~K$\alpha$ line, failed to provide a significant improvement to the fit. As an additional test, we used the \XMM pn data to model emission in the Fe K bandpass. The best-fitting energy of the Fe~K$\alpha$ line in the \XMM data is $6.41\pm0.02$ keV, and for the Fe~K$\beta$ line is $7.05\pm0.06$ keV. The ratio of the intensities of the two lines is consistent with the canonical value 150/17.

Next, we measured the strong emission from Si K$\alpha$ that is detected in the \Ch HETGS spectra. We used the good effective area of both the \Ch MEG and HEG instruments at 1.7--2 keV to search for emission and absorption features associated with Si. The data were grouped to a minimum of 4 channels per bin with at least 20 counts in each bin. We used the initial continuum model described in Section~\ref{initialspectralfitting} plus a Si~K edge (discussed in detail in Section~\ref{sikedge}) to fit the spectra over an energy range 1.5--6 keV (MEG) and 1.5--9 keV (HEG), excluding the Fe~K bandpass of 6.0--7.2 keV and, initially, the Si K$\alpha$ bandpass of 1.70--1.76 keV. This model provided a good fit to the data ($\chi^2=835.22$ for 831 dof). We included the energy range 1.70--1.76 keV in our spectral fitting, and still found an acceptable fit to the data ($\chi^2=905.56$ for 868 dof). Allowing the continuum parameters to vary, we added a Gaussian of initial centroid energy of 1.74~keV in order to model the Si~K$\alpha$ emission line, and allowed its energy, width and normalization to be free parameters. The inclusion of the line resulted in a significant decrease in the fitting statistic to $\chi^2=873.35$ for 865 dof. Figure~\ref{newgp_try3} shows the MEG spectrum and best-fitting model in the Si K bandpass. Figure~\ref{newgp_try3_contours} shows confidence contours of the energy and width of the Si K$\alpha$ line. The line is marginally resolved, with a 90\% confidence (for two interesting parameters) width of $\sigma=3.6^{+2.4}_{-2.0}$ eV [c.f., the instrumental resolution at this energy of $\sigma\simeq$1 eV (HEG) $\sigma\simeq$2 eV (MEG)]. This corresponds to a velocity width of $600^{+400}_{-300}$ km s$^{-1}$. We note that $\chi^2=884.81$ for 866 dof when the line width was frozen at an unresolved value. The energy of the line, $E=1.740\pm0.002$ keV (90\% confidence for two interesting parameters), is consistent with emission from neutral silicon. The equivalent width of the line is $5.6^{+2.0}_{-1.8}$~eV.

In addition to the Si K$\alpha$ line, we report the detection of S K$\alpha$ emission, as well as tentative detections of Ar K$\alpha$ and Ca K$\alpha$ lines. The parameters of all fluorescent emission lines detected with the \Ch HETGS are given in Table~\ref{kalphalines}.

\subsection{Evidence for modest broadening?}
\label{evidenceforbroadening}

Figure~\ref{newfe_oplot} shows the HEG counts spectrum in the energy range 5.8--7 keV, together with the model fit consisting of a continuum and narrow Fe~K$\alpha$ line described in Section~\ref{fe_si_narrowcores} (shown in light gray). There appears to be evidence for modest broadened emission in the few bins either side of the base of the narrow Fe~K$\alpha$ line core. We added to the model a second, more broadened, Gaussian line, with its energy tied to that of the narrow core and left its width and normalization as free parameters. This resulted in an improved fit to the data ($\chi^2=162.89$ for 152 dof), which is significant at 95\% on an $F$-test. The best-fitting width of this broadened feature is $\sigma=96^{+71}_{-47}$~eV, which corresponds to a velocity width of $4500^{+3300}_{-2200}$~km~s$^{-1}$. The equivalent width of the broad feature in this model is $63^{+39}_{-40}$~eV, and is $81^{+27}_{-30}$~eV for the narrow core (see Figure~\ref{newfe_oplot}).

\section{Excess Absorption}
\label{absorptionfeatures}
\subsection{The Fe K Edge}

We used the high signal-to-noise of the \XMM pn observation to determine if any excess absorption at the Fe~K edge is present, as first detected by \cite{mal99} using {\it BeppoSAX}. We prefer to use the \XMM data, owing to their higher S/N when compared with the \Ch HETGS data. The model fits were performed between 4 and 9 keV, and the data were grouped to 100 counts per bin. We fitted a model consisting of an absorbed power law and two narrow Gaussian emission lines (with $\sigma$ frozen at 10 eV) to model contributions from Fe~K$\alpha$ and K$\beta$, which provided an acceptable fit ($\chi^2=188.62$ for 195 dof). We added an additional neutral Fe~K edge to the model, which resulted in an improved fit to the spectrum ($\chi^2=183.62$ for 193 dof). The best-fitting optical depth of the edge is $0.07^{+0.06}_{-0.04}$, which is consistent with the value found by \cite{mal99}, and corresponds to a column density $N_{\rm H}=(6^{+5}_{-3})\times10^{22}$ cm$^{-2}$ assuming the Fe cross section given by \cite{lea93}. Finally, we searched for additional Fe~K absorption in the \Ch data, and measured an excess optical depth that is consistent with the \XMM data.

\subsection{The Si K Edge}
\label{sikedge}

We found evidence for excess absorption at Si K in the \Ch HETGS data. The HEG and MEG spectra were grouped to a minimum of 4 channels per bin with at least 20 counts in each bin. We added an additional edge to the continuum model found in Section~\ref{initialspectralfitting}, and allowed the continuum parameters to vary. This resulted in a large improvement in the fit ($\Delta\chi^2=22.72$ for two additional parameters). The best-fitting energy of the edge is $1.847^{+0.004}_{-0.006}$ keV and its optical depth $0.13^{+0.05}_{-0.04}$. The energy of the edge is consistent with neutral or near-neutral states of Si (the neutral Si~K edge has an energy of 1.839 keV). Calibration uncertainties of the MEG and HEG instruments introduce an additional systematic error to the optical depth of $\pm0.04$ (M. Nowak, private communication). As an additional test, we searched for excess Si~K absorption in the \XMM data, and found an excess optical depth of $\tau=0.12\pm0.04$, consistent with the \Ch data.

\section{Evidence for Multiple layers of Absorption and the final parametrization of the broad-band spectrum}
\label{finalmodel}

We have demonstrated that there is evidence for absorption in the nuclear spectrum of NGC~2110 that cannot be fully accounted for by a fully-covering, Compton-thin ($N_{\rm H}\sim3\times10^{22}$ cm$^{-2}$) absorber. In our present parametrization of the continuum (see Section~\ref{initialspectralfitting}), the photon index of the continuum is unusually flat ($\Gamma\sim1.4$), although {\it BeppoSAX} observations suggest that the photon index steepens to $\Gamma$=1.9 when data above 13~keV are considered (\citealt{mal99}). Motivated by this, we investigated the possibility that an {\it additional} layer of cold absorption is required to adequately fit the 0.5--9~keV spectrum. We adopted a partial-covering model, implemented in {\sc XSPEC/ISIS} as \textsc{zwabs$\times$zpcfabs$\times$zpcfabs(powerlaw)}. The individual column densities $N_{\rm H, 1}$, $N_{\rm H, 2}$, and $N_{\rm H, 3}$, respectively have covering fractions $c_1$, $c_2$, and $c_3$, where $c_3$=1 is a fully-covering screen (in order to model very mild absorption associated with the soft excess that we detected --- see Section~\ref{initialspectralfitting}). We included neutral fluorescent lines from Fe and Si, and allowed the normalizations of all components to vary across the three datasets. The data were grouped to 100 counts per bin, and the model fits were performed between 0.5--9~keV, with the exception of the MEG, for which the model fits were performed between 0.5--6~keV.

The best-fitting model consisted of a power law of photon index $\Gamma\sim1.7$, absorbed by columns $N_{\rm H, 1}$=$1.3\times10^{23}$~cm$^{-2}$ ($c_1$=$32\pm1$\%), $N_{\rm H, 2}$=$2.8\times10^{22}$~cm$^{-2}$ ($c_2$=$96\pm0.1$\%), and $N_{\rm H, 3}$=$7.7\times10^{20}$~cm$^{-2}$ ($c_3$=$100$\%). There are significiant degeneracies between the covering fractions, and so we simply quote the 90\% confidence (for one interesting parameter) uncertainties on these parameters. In terms of the transmission fractions, $\sim$31\% of the emission is absorbed by an integrated column of $1.6\times10^{23}$~cm$^{-2}$, $\sim$1\% by $12.9\times10^{23}$~cm$^{-2}$, $\sim$65\% by $2.8\times10^{22}$~cm$^{-2}$, and the remaining 3\% by $7.7\times10^{20}$~cm$^{-2}$. This model provided an excellent fit to the spectrum ($\chi^2$=828.84 for 993 dof), which is a substantial improvement over the model described in Section~\ref{initialspectralfitting} ($\chi^2$=881.14 for 995 dof). Further, the photon index of the power law is significantly steeper than the previous model, and is now consistent with typical values found in Seyfert galaxies. The inclusion of the $8\times10^{20}$~cm$^{-2}$ absorber provides a significantly better fit to the data over models in which this extra absorption is absent. This column may represent an additional Galactic column towards NGC~2110, since the value we adopted in our analysis ($N_{\rm H, Gal}$ = 1.76$\times$10\(^{21}\) cm$^{-2}$ --- \citealt{dic90}) is an interpolated one. The parameters of the best-fitting multiple absorber model are given in Table~\ref{multipleabsorber_tab}, and the MEG, HEG, and \XMM pn counts spectra are shown in Figure~\ref{multipleabsorber}.

Given that there is clear evidence of Compton reflection as suggested by the strong detections of fluorescent K$\alpha$ line emission, we evaluate the reflection spectrum by including a continuum in the form of a {\sc pexrav} model (\citealt{mz95}). We added to the spectral model the reflection-only component of the {\sc pexrav} model, with the photon index tied to that of the primary power law, the inclination angle frozen at 45$^{\circ}$, abundances frozen at solar, and free normalization (although we linked the normalizations of the MEG and HEG). The addition of the {\sc pexrav} component resulted in a modest decrease in the fitting statistic ($\chi^2$=824.00 for 991 dof), which highlights the relative insensitivity of the data to the reflection component at energies $<$10~keV.

Finally, we investigated the Fe~K$\alpha$ line profile with the new continuum models and found the addition of more complex absorption or reflection components had no effect on the measured parameters of the narrow line core or modestly broadened feature, which retrospectively justifies our use of a simple, absorbed power law model for measuring the line parameters in Section~\ref{reflectionspectrum}.

\section{The Low-Energy Spectrum}
\label{lowenergyspectrum}

We examined data at energies $<$1 keV from the \Ch MEG and \XMM RGS instruments. The detection of features in this energy range is difficult, owing to the high Galactic column towards NGC~2110, as well as the modest intrinsic flux of the source at this energy. To make a better assessment of genuine emission and absorption lines in the spectrum, we considered data from the two \Ch epochs separately, and did not co-add the data from the two \XMM RGS cameras.

All data were binned to the full instrumental resolution of the \XMM RGS camera at 0.5 keV of $\sim$0.075\AA. A spurious emission feature was detected in the \XMM RGS1 camera at an observed energy of $\sim$0.37 keV. This feature was not evident in the RGS2 camera, and is possibly associated with anomalously high dark currents in CCD2\footnote{See http://xmm.vilspa.esa.es/external/xmm\_user\_support/ documentation/uhb/XMM\_UHB.html}. The four spectra in the energy range 0.5--1 keV are shown in Figure~\ref{4panel_lowe}.

We searched for features in the low energy spectrum by plotting all potential (1) K$\alpha$ transitions from neutral elements, (2) H- and He-like emission lines, (3) radiative recombination continuua (RRC) features, and (4) other 0.5--1 keV transitions detected with \XMM and \Ch grating observations of the Seyfert 2 galaxy, NGC~1068 (\citealt{kin02,ogle03}). There appears to be tentative evidence of ionized emission in the form of O~{\sc viii}~Ly~$\alpha$ emission in three of the four observations, as well as the O~{\sc viii}~RRC feature first reported by \cite{gua06}. A potential Ne~{\sc ix} forbidden ({\it z}) line is also observed at the 1--2$\sigma$ level. There appears to be a line in the spectra of both RGS instruments at an observed energy of 0.76~keV. However, this line is not detected in either \Ch MEG observation, it does not correspond to any of the transitions we searched for above, and its energy is inconsistent with any of the prominent lines in the collisionally ionized plasmas used by \cite{evans06} to model the extended X-ray gas environment of the source. Therefore, we do not study this feature further, and speculate that it may be an instrumental artifact.

The potential detection of such ionized features, even at modest significance, is interesting, and we discuss this further in Section~\ref{interp}.

\section{CCD spectroscopy of the contaminating point source}
\label{unrelatedps}

In our initial data analysis, we identified a contaminating soft X-ray source coincident with the MEG+1 dispersion axis in the 2003 March \Ch observation (OBSID 4377). The source lies $\sim9'$ from the nucleus, at a position 05:52:03.145, -07:18:40.65 (J2000), and is marginally spatially extended in our \Ch image. Inspection of an archival 2MASS image shows a source of J-band magnitude 9.2 that is spatially coincident with the soft X-ray source, and is accompanied by a fainter source of J-band magnitude 12.0 that lies $\sim7''$ southeast and extends along the same position angle as the \Ch source. 

We extracted the X-ray spectrum of the source(s) using the {\sc CIAO} {\sc specextract} script from a $24''\times14''$ rectangle, with background sampled from a large off-source region free from point sources. Single-component models of either a power law or thermal {\sc apec} emission of abundance fixed at solar failed to provide an adequate fit to the spectrum, but the combination of the two provided a good fit to the data ($\chi^2=21.96$ for 15 dof). The best-fitting parameters are $\Gamma=2.6\pm0.4$ and $kT=1.0\pm0.2$ keV, and the 0.5--2 keV unabsorbed X-ray flux is $3.6\times10^{-13}$ ergs cm$^{-2}$ s$^{-1}$.

The origin of the source(s) is unclear, although we note that its X-ray spectrum might be consistent with an AGN embedded in a group-scale X-ray gas atmosphere.

\section{Interpretation}
\label{interp}

The nuclear spectrum of NGC~2110 is well-described by a power law that is absorbed through one or more layers of neutral material, together with a reflection spectrum in the form of a fluorescent K$\alpha$ lines from Si, S, Ar, Ca, and Fe. In this section, we use the K$\alpha$ lines to diagnose the origin of the fluorescent emission (\S\ref{interp-kalpha}), discuss alternative models for the geometry of the circumnuclear absorption (\S\ref{interp-circumnuclear}), consider NGC~2110 in the context of the Narrow Emission Line Galaxy phenomenon (\S\ref{interp-nelg}), and finally (\S\ref{interp-soft}) comment on the physical processes responsible for the soft X-ray emission.

\subsection{Fe K$\alpha$ line diagnostics}
\label{interp-kalpha}

The strong Si~K$\alpha$ and Fe~K$\alpha$ lines detected with the \Ch HETGS allow us to use them to place constraints on the location and physical state of the fluorescing material. The energies of the line cores, $1.740\pm0.002$ keV and $6.397\pm0.007$ keV respectively, are consistent with fluorescence from neutral or near-neutral species of Si and Fe. The measured K$\alpha$ line width and the estimated black hole mass of $2\times10^{8}$ M$_\odot$ (\citealt{woo02}) allow us to constrain the emission radius to be $\gappeq1$~pc, using Keplerian arguments. In short, the reflection spectrum of the nucleus of NGC~2110 is dominated by fluorescent emission from neutral material at a large distance from the central supermassive black hole

The \Ch and \XMM data presented here allow us to comment on previous attempts with {\it ASCA} to model the Fe~K$\alpha$ line complex in NGC~2110 (as part of a co-added sample of Seyfert 2 spectra) as having a significant contribution from a relativistically blurred ``diskline'', either oriented at an intermediate angle to the line of sight (\citealt{wea98}) or close to face on (\citealt{tur98}). The data rule out the presence of a strong diskline component that was detected down to $\sim$~4~keV in the co-added sample of \cite{tur98}, and also highlight the insensitivity of {\it ASCA} to the complex layers of absorption we detected with \Ch and \XMM in NGC~2110. There is tentative evidence for a slightly broadened ($4500^{+3300}_{-2200}$ km~s$^{-1}$) component to the line profile, but this cannot be construed to be diskline emission. Unmodeled absorption could explain the presence of the red wing of the Fe~K$\alpha$ profile in the {\it ASCA} sample, as has been suggested for NGC~3783 (\citealt{reeves04}) and NGC~3516 (\citealt{tur05}), and we note in general that it is of critical importance to accurately model absorption effects in order to deconvolve the iron line complex from the underlying continuum.

The lack of highly broadened Fe~K$\alpha$ line emission may imply that reflection from the very innermost portions of an accretion disk is genuinely absent from NGC~2110; further support for this model is confirmed by the substantially sub-Eddington luminosity of the accretion flow ($L_{\rm X}$/$L_{\rm Edd}$$\sim$$10^{-3}$, or $L_{\rm bol}$/$L_{\rm Edd}$$\sim$$10^{-2}$ if the X-ray luminosity is 10\% of the bolometric luminosity), and by new observations of the $>$10 keV spectrum with {\it Suzaku} that place a strong upper limit to the reflection fraction, $R$, of $<$10\% (\citealt{reeves06}, Okajima et al. 2007, in preparation).

\subsection{The origin of the circumnuclear absorption}
\label{interp-circumnuclear}

A naive initial parametrization of the nuclear spectrum of NGC~2110 as the sum of (1) a $\Gamma\sim1.4$ power law, absorbed by a column of $2.8\times10^{22}$ cm$^{-2}$, and (2) a lightly absorbed power law of identical photon index, has several difficulties in accounting for the detailed properties of the spectrum. First, the spectrum has significantly larger Si and Fe K edges that could be accounted for by the initial model. Second, the Fe~K$\alpha$ equivalent width of $80\pm30$~eV is rather larger than the $\sim$30--40~eV that would be predicted from an absorbing column of $3\times10^{22}$ cm$^{-2}$ (e.g.,~\citealt{miy96}). Finally, the X-ray spectrum of the source is known to steepen to $\Gamma$$\sim$$1.9\pm0.1$ when data above 13 keV are considered (\citealt{mal99}). These pieces of evidence indicate that a more complex spectral model is necessary to adequately model the spectrum.

There are several models that may account for the nuclear spectrum. We considered in detail a partially covered power law of photon index $\Gamma\sim1.7$, absorbed by columns $N_{\rm H, 1}$=$1.3\times10^{23}$~cm$^{-2}$ (with a covering fraction of 32\%), $N_{\rm H, 2}$=$2.8\times10^{22}$~cm$^{-2}$ (covering 96\% of the nuclear emission), together with a fully-covering screen of $N_{\rm H, 3}$=$7.7\times10^{20}$~cm$^{-2}$. In terms of the transmission fractions, $\sim$31\% of the emission is absorbed by an integrated column of $1.6\times10^{23}$~cm$^{-2}$, $\sim$1\% by $12.9\times10^{23}$~cm$^{-2}$, $\sim$65\% by $2.8\times10^{22}$~cm$^{-2}$, and the remaining 3\% by $7.7\times10^{20}$~cm$^{-2}$. The addition of the $10^{23}$~cm$^{-2}$ coverer produces sufficient opacity at the Si and Fe K edges to be consistent with the data, as discussed by \cite{hay96}, and the predicted equivalent width is consistent with the measured value. Further, this model provides an elegant means of reconciling the 2--10 keV power law photon index with that of the $>$10~keV spectrum. Although we cannot make a definitive claim as to the origin of the partially covering $10^{23}$~cm$^{-2}$ absorber, it remains plausible that it originates in broad-line region clouds in the vicinity of the AGN, as first proposed by \cite{hay96}. The $3\times10^{22}$~cm$^{-2}$ absorber through which 65\% of the nuclear emission is transmitted is likely to (1) have a flattened geometry in order to allow the small scale radio jet to escape and (2) be at a distance further from the supermassive black hole than the majority of the optical broad line region. The required large distance is consistent with the measured velocity width of the narrow Fe~K$\alpha$ line core.

Alternatively, the observed spectrum could be reproduced with a model in which the circumnuclear absorbing structure is partially ionized. With a more modest column than the neutral partial coverer model described above, a partially ionized absorber would transmit some low-energy continuum emission, and one could instead associate resonance absorption effects from a partially ionized coverer between 7--8 keV (e.g.,~\citealt{kal04}) with the K edge of neutral Fe in low signal-to-noise data. However, our \Ch and \XMM observations are insufficient to test the broad-band properties of this model, since a measurement of the high-energy continuum is necessary to constrain the photon index of the power law. However, we find no evidence of discrete ionized absorption features in the spectrum, which may disfavor this model. Observations with {\it Suzaku} (Okajima et al., in preparation) will be able to better evaluate the ionized absorber model.

\subsection{NGC 2110 in the context of Seyfert and Narrow Emission Line Galaxies}
\label{interp-nelg}

NGC~2110 was originally classed as a Narrow Emission Line Galaxy due to it having the Seyfert 1-like characteristics of strong flat-spectrum ($\Gamma$$\sim$1.5) 2--10 keV X-ray emission and a broad H$\alpha$ line, but Seyfert 2-like features in the form of prominent narrow ($<600$km s$^{-1}$) optical emission lines and significant X-ray absorption (\citealt{shu80,mus82}). This led \cite{law82} to suggest that NELGs may represent transition objects between the two classes.

Our new \Ch and \XMM observations have shown that the power-law photon index is {\it not} flat, once the complex absorption of the nucleus is taken into account, and we suggest that the X-ray spectrum of NGC~2110 is consistent with that of a Seyfert 2 galaxy. We speculate that the broad H$\alpha$ line could be caused by transmission through the ``leaky'' absorber, since a small fraction of the nuclear emission escapes through a very modest column. Although we cannot place constraints on the inclination of the nucleus with respect to the observer by modeling diskline emission, we note that (1) the high intrinsic absorption of the source and (2) the two-sided radio jet both suggest that the nucleus is inclined at low (i.e., edge-on) to intermediate angles, which would be consistent with AGN unification schemes.

\subsection{The soft excess and ionized emission lines}
\label{interp-soft}

The large Galactic column towards NGC~2110 means that detections of features in the low-energy gratings spectra are at best tentative. However, the possible detections of O~{\sc viii}~Ly~$\alpha$ emission and a narrow O~{\sc viii}~RRC feature (first reported by \citealt{gua06}) allow us to place some constraints on the physical processes associated with the soft excess. We (\citealt{evans06}) detected resolved X-ray emission that lies $\sim$160~pc north of the nucleus, and considered several models that may account for it, including collisional ionization, photoionization, and electron-scattering of nuclear radiation. The potential detection of the narrow O~{\sc viii}~RRC feature may suggest that photoionization processes are important in the vicinity of the nucleus of NGC~2110, though we cannot rule out collisional ionization being responsible for the extended X-ray circumnuclear environment.

\section{CONCLUSIONS}

We have presented results from {\it Chandra} HETGS and {\it XMM-Newton} EPIC and RGS observations of NGC~2110, a source historically classified as a nearby Narrow Emission Line Galaxy. Our results support the interpretation that the X-ray spectrum of the source is dominated by absorption and fluorescence from neutral material at a large distance from the central supermassive black hole, and that the source is consistent with a Seyfert 2 galaxy viewed through a patchy absorber. To summarize our conclusions:

\begin{enumerate}

\item The continuum X-ray spectrum of the source is best described by a power law of photon index $\Gamma\sim1.7$, absorbed by columns $N_{\rm H, 1}$=$1.3\times10^{23}$~cm$^{-2}$ (with a covering fraction of 32\%), $N_{\rm H, 2}$=$2.8\times10^{22}$~cm$^{-2}$ (covering 96\%), together with a fully-covering screen of $N_{\rm H, 3}$=$7.7\times10^{20}$~cm$^{-2}$. In terms of the transmission fractions, $\sim$31\% of the emission is absorbed by an integrated column of $1.6\times10^{23}$~cm$^{-2}$, $\sim$1\% by $12.9\times10^{23}$~cm$^{-2}$, $\sim$65\% by $2.8\times10^{22}$~cm$^{-2}$, and the remaining 3\% by $7.7\times10^{20}$~cm$^{-2}$.

\item This multiple partial coverer model has a number of advantages over a more simple parametrization of the X-ray spectrum. First, the photon index of the power law is significantly steeper than in simple models, which reconciles the shape of the 2--10 keV continuum with that of the $>$10~keV spectrum and demonstrates that this NELG does not have a flat photon index as first suggested. Second, the model provides sufficient opacity at the neutral Si and Fe K edges to be consistent with the spectrum.

\item We detect marginally resolved emission from neutral Fe~K$\alpha$ and Si~K$\alpha$ lines with the \Ch HETGS, which allows us to constrain the emission radius of these fluorescent features to $\gappeq1$ pc from the central supermassive black hole. The data rule out the presence of a strong diskline, and we suggest that inner-disk reflection is absent in NGC~2110.

\item We report the additional detection of fluorescent S~K$\alpha$ emission, as well as tentative Ar and Ca K$\alpha$ lines.

\item The tentative detections of ionized emission lines in the low-energy gratings spectra may support models in which collisional ionization or photoionization are important processes in the vicinity of the NGC~2110 nucleus.

\item We suggest that the $10^{23}$~cm$^{-2}$ partially covering absorber takes the form of broad-line region clouds in the vicinity of the AGN, and that the $3\times10^{22}$~cm$^{-2}$ coverer is likely to have a flattened geometry in order to allow the small-scale radio jet to escape and is at a significant distance from the supermassive black hole for it to obscure the majority of the optical broad line region.

\item The measured photon index and high intrinsic absorption of the nuclear spectrum of NGC~2110 are consistent with that of a typical Seyfert 2 galaxy. This, together with the two-sided radio jet, may imply that the nucleus is oriented at edge-on to intermediate angles to the line of sight, consistent with unified models.

\end{enumerate}

\acknowledgements

This work was supported from the National Aeronautics and Space Administration through Chandra Award Number GO2-3156X, issued by the Chandra X-ray Observatory Center, which is operated by the Smithsonian Astrophysical Observatory for and on behalf of the National Aeronautics and Space Administration under contract NAS8-03060. We thank James Reeves for detailed discussions of the X-ray properties of NGC~2110, and Ralph Kraft, Martin Elvis, and Mike Nowak for their comments. We wish to thank the anonymous referee for very useful and constructive comments. This research has made use of the NASA/IPAC Extragalactic Database (NED) which is operated by the Jet Propulsion Laboratory, California Institute of Technology, under contract with the National Aeronautics and Space Administration, and of the SIMBAD database, operated at CDS, Strasbourg, France.

\begin{figure}
\begin{center}
\includegraphics[width=8cm]{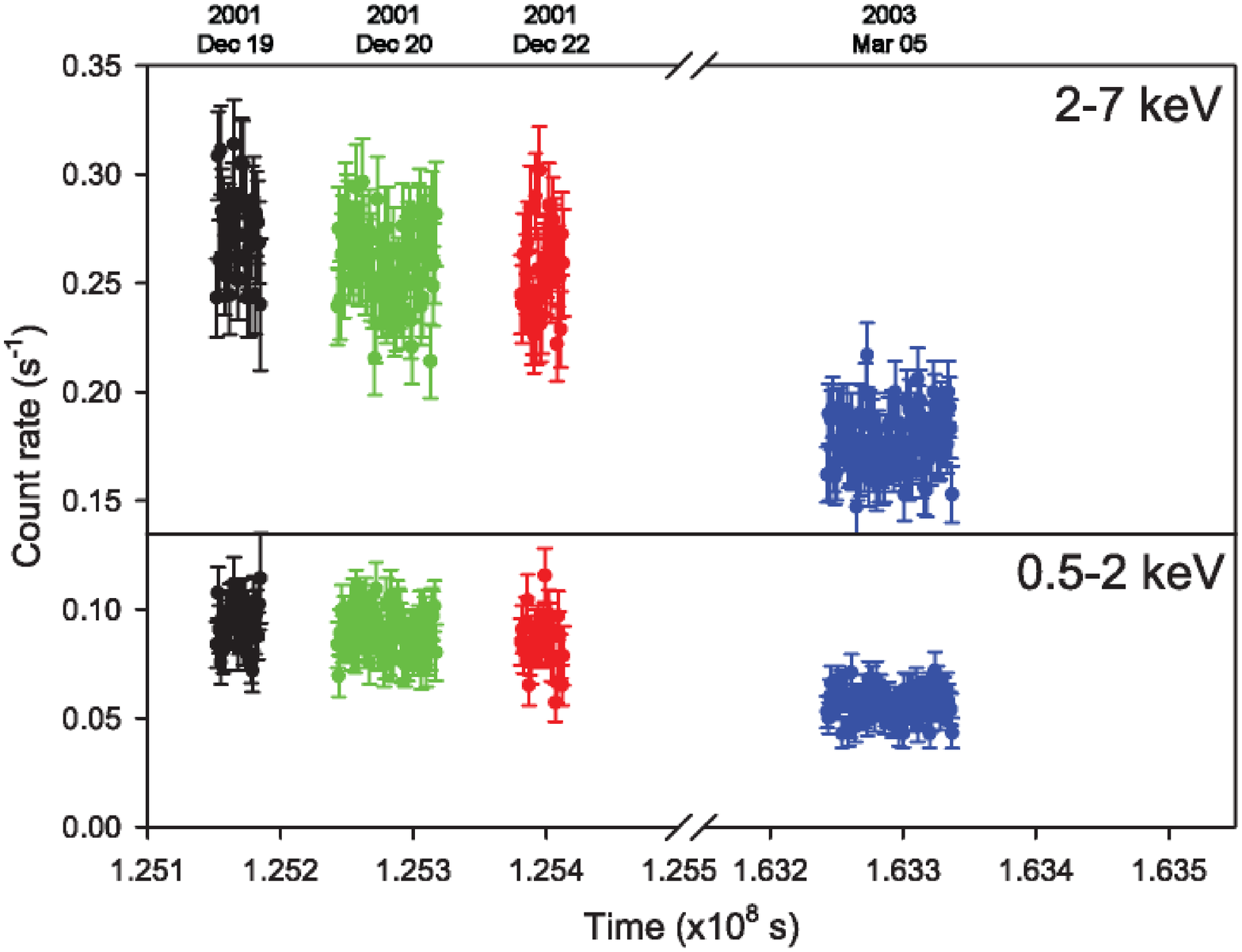}
\caption{Lightcurves for the four epochs of \Ch HETGS data, in the energy ranges 0.5--2 keV and 2--7 keV, and binned to 500-second time intervals. The first three observations, performed in 2001 December, show no variability. However, the fourth observation, performed in 2003 March, shows a significant decrease in count rate. Here, black points correspond to data from OBSID 3143, green to OBSID 3418, red to OBSID 3417, and blue to OBSID 4377.}
\label{hetgs_lightcurves}
\end{center}
\end{figure}

\begin{figure}
\begin{center}
\includegraphics[width=8cm,angle=0]{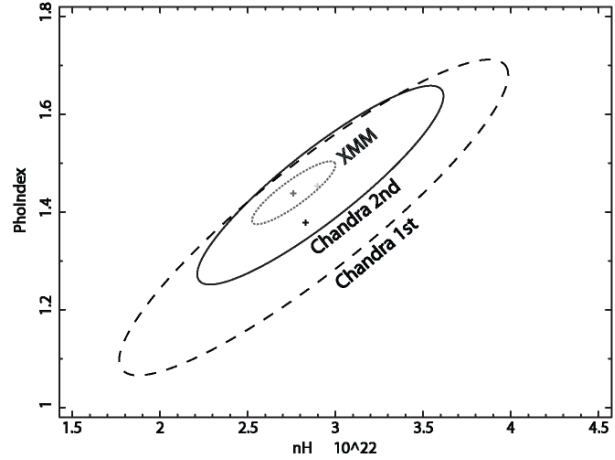}
\caption{90\% confidence (for two interesting parameters) contours of the intrinsic absorption and photon index for the 2001 December \Ch HETGS observations (OBSIDs 3143, 3417, and 3418 combined) ({\it black solid line}), the 2003 March \Ch HETGS observation ({\it black long-dashed lines}), and the 2003 March \XMM EPIC observation ({\it dark gray short-dashed line}).}
\label{contours_lightcurves}
\end{center}
\end{figure}

\begin{figure}
\begin{center}
\includegraphics[height=8cm,angle=270]{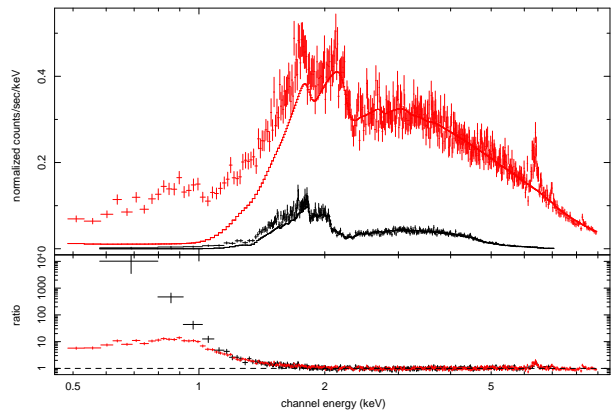}
\caption{0.5--9 keV \Ch MEG ({\it black}) and \XMM pn ({\it red}) spectra and data/model ratio of NGC~2110, plotted against an absorbed power law fitted in the 2--9 keV band. A strong soft excess is seen at energies $\lappeq1.5$ keV, and emission in the Fe~K$\alpha$ bandpass (not included in the continuum fitting) is also observed.}
\label{softexcess_merged_MEG_pn_try1}
\end{center}
\end{figure}

\begin{figure}
\begin{center}
\includegraphics[height=8cm,angle=270]{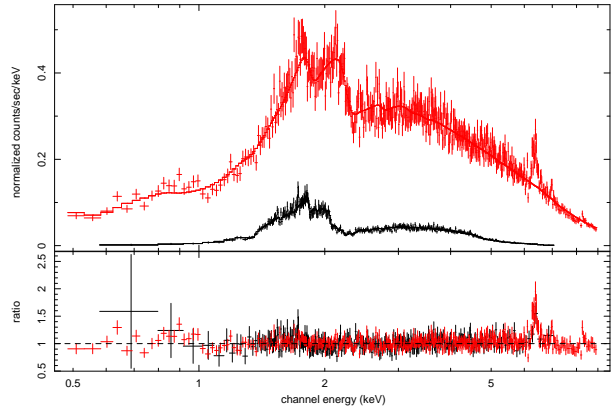}
\caption{0.5--9 keV \Ch MEG ({\it black}) and \XMM pn ({\it red}) spectra and data/model ratio of NGC~2110, plotted against a model consisting of the sum of a heavily absorbed power law and a mildly absorbed power law.}
\label{softexcess_merged_MEG_pn_try3}
\end{center}
\end{figure}

\begin{figure}
\begin{center}
\includegraphics[width=8cm,angle=0]{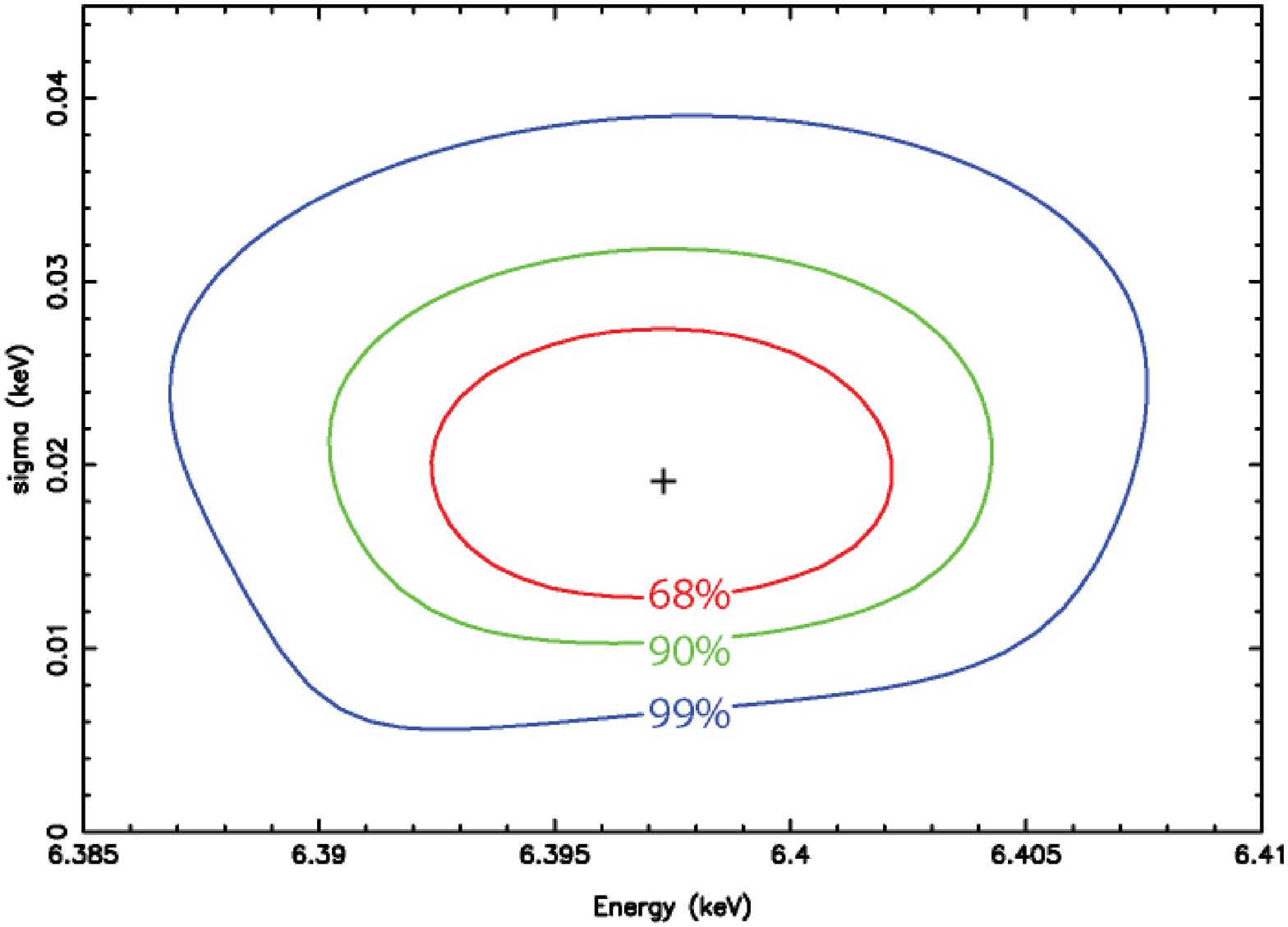}
\caption{Energy and linewidth confidence contours for the Fe~K$\alpha$ line core measured from \Ch HEG spectrum. Shown are the 68\%, 90\%, and 99\% contours.}
\label{newfe_try2_contours}
\end{center}
\end{figure}

\begin{figure}
\begin{center}
\includegraphics[width=8cm,angle=0]{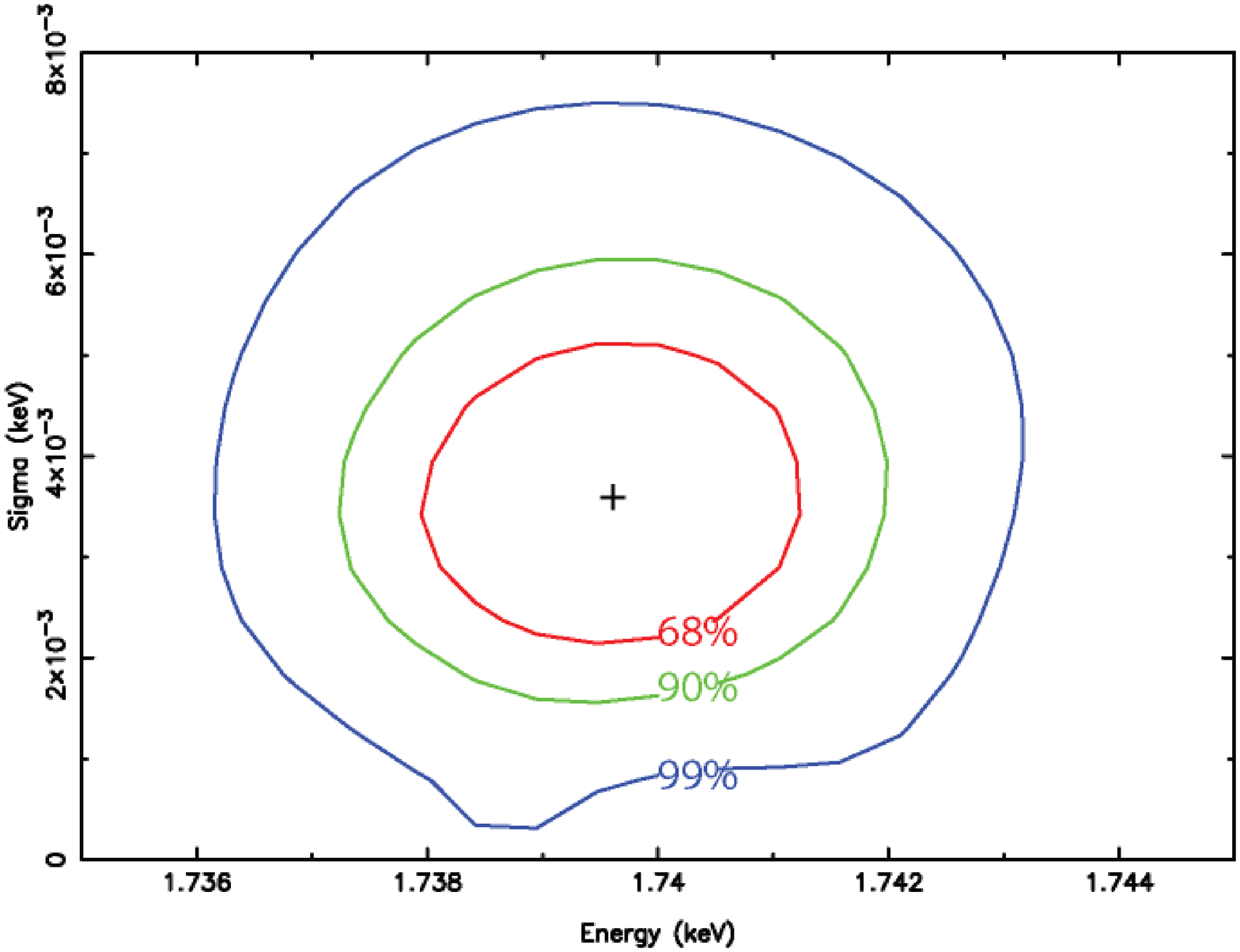}
\caption{Energy and linewidth confidence contours for the Si K$\alpha$ line core measured from the \Ch HETGS spectra. Shown are the 68\%, 90\%, and 99\% contours.}
\label{newgp_try3_contours}
\end{center}
\end{figure}

\begin{figure}
\begin{center}
\includegraphics[width=8cm,angle=0]{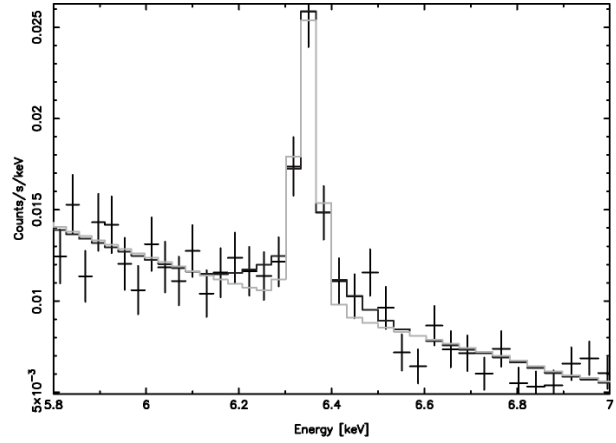}
\caption{Full-resolution \Ch HEG counts spectrum in the Fe~K$\alpha$ bandpass, showing model fits to the continuum plus: ({\it light gray}) a narrow Gaussian Fe~K$\alpha$ line core; ({\it black}) a narrow Gaussian Fe~K$\alpha$ line core plus a broad Gaussian Fe~K$\alpha$ base.}
\label{newfe_oplot}
\end{center}
\end{figure}

\begin{figure}
\begin{center}
\includegraphics[height=8cm,angle=270]{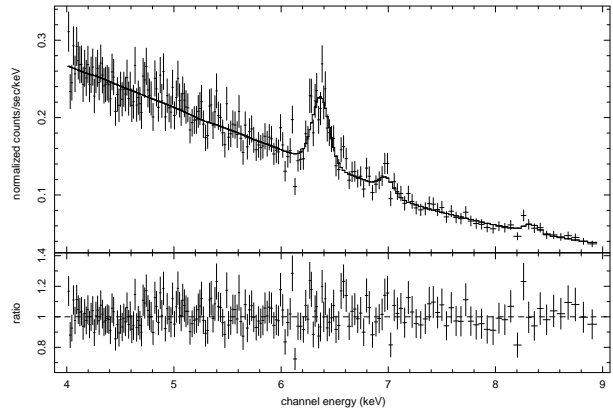}
\caption{\XMM pn counts spectrum and data/model residuals in the Fe~K$\alpha$ bandpass. The model fit shown is the sum of an absorbed power law, an additional Fe~K edge and Fe~K$\alpha$ and K$\beta$ emission lines.}
\label{ralph1}
\end{center}
\end{figure}

\begin{figure}
\begin{center}
\includegraphics[height=8cm,angle=270]{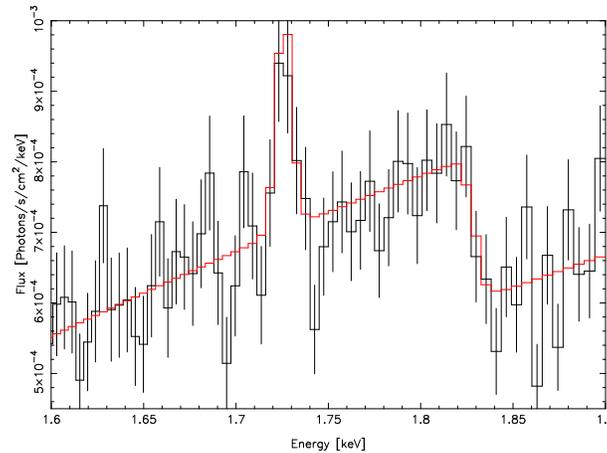}
\caption{\Ch MEG flux spectrum in the energy range 1.6--1.9 keV. The model fit is the sum of a heavily absorbed power law and a lightly absorbed power law, an additional Si K edge and emission from neutral Si K$\alpha$.}
\label{newgp_try3}
\end{center}
\end{figure}

\begin{figure}
\begin{center}
\includegraphics[height=8cm,angle=270]{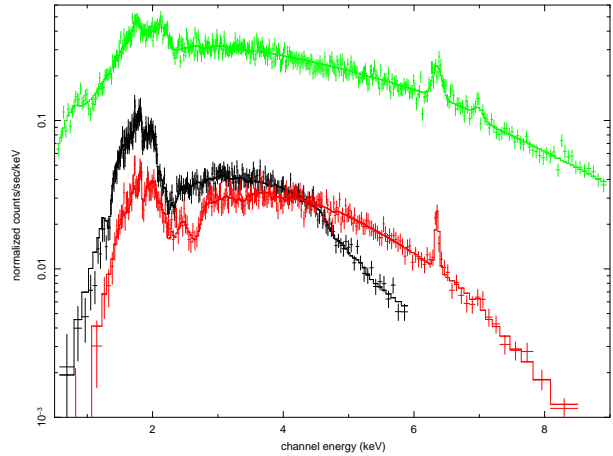}
\caption{0.5--9 keV \Ch MEG ({\it black}), \Ch HEG ({\it red}) and \XMM pn ({\it green}) counts spectra of NGC~2110, plotted against the final model of a power law of photon index $\Gamma\sim1.7$ partially covered by a multiple layers of absorption. Neutral fluorescent lines from Si and Fe are also modeled.}
\label{multipleabsorber}
\end{center}
\end{figure}

\begin{figure}
\begin{center}
\includegraphics[width=17cm,angle=0]{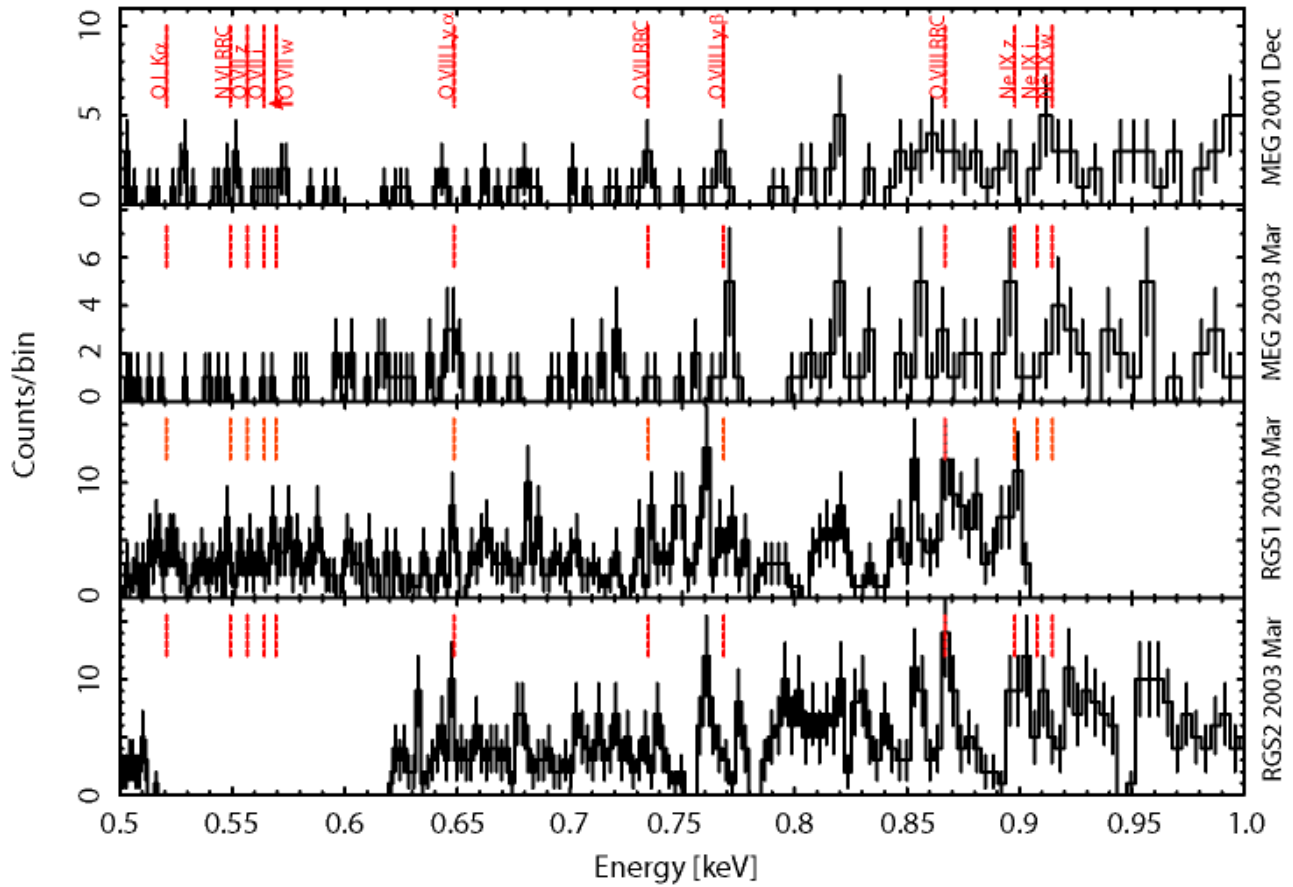}
\caption{0.5--1 keV spectra of NGC~2110, using data from \Ch MEG data in 2001 December ({\it top panel}, \Ch MEG data in 2003 March ({\it second panel}, \XMM RGS1 data in 2003 March ({\it third panel}, and \XMM RGS2 data in 2003 March ({\it bottom panel}. Plotted are the transitions that are tentatively detected; O~{\sc viii}~Ly~$\alpha$ emission, an O~{\sc viii}~RRC, and a forbidden ({\it z}) Ne~{\sc ix} line. For comparison, we also show O~{\sc viii}~Ly~$\beta$ emission, the O~{\sc vii}~RRC, the O~{\sc vii} and Ne~{\sc ix} triplets, and neutral O~K$\alpha$ line. We caution that any detection of features in this low-energy spectrum is at best marginal.}
\label{4panel_lowe}
\end{center}
\end{figure}

\clearpage
\begin{deluxetable}{llllc}
\tablecaption{Observation Log}
\tablehead{Telescope & Instrument & Date & Observation ID & Nominal Exposure (ks)}
\startdata
\Ch   & HETGS   & 2001 December 19 & 3143        & 35       \\
\Ch   & HETGS   & 2001 December 20 & 3418        & 80       \\
\Ch   & HETGS   & 2001 December 22 & 3417        & 35       \\
\Ch   & HETGS   & 2003 March 05    & 4377        & 100      \\
\XMM  & EPIC    & 2003 March 05    & 0145670101  & 60       \\
\XMM  & RGS     & 2003 March 05    & 0145670101  & 60       
\enddata
\label{obslog}
\end{deluxetable}

\begin{deluxetable}{llll}
\tablecaption{Results from initial spectral fitting to the broad-band continuum}
\tablehead{
Component  & Parameters    & Normalization & 0.5--10 keV unabsorbed \\
           &               &               & luminosity (ergs s$^{-1}$)}
\startdata
Soft PL    & $N_{\rm H}=(4.16^{+2.37}_{-2.29})\times10^{20}$ cm$^{-2}$ & MEG=$(2.66\pm0.42)\times10^{-4}$          & $(3.34\pm0.53)\times10^{41}$ \\
           & $\Gamma=1.40\pm0.04$                                      & XMM=$(2.47^{+0.23}_{-0.21})\times10^{-4}$ & $(3.10^{+0.29}_{-0.26})\times10^{41}$ \\ \hline
Hard PL    & $N_{\rm H}=(2.80\pm0.10)\times10^{22}$ cm$^{-2}$          & MEG=$(6.49^{+0.37}_{-0.34})\times10^{-3}$ & $(8.15^{+0.46}_{-0.43})\times10^{42}$ \\
           & $\Gamma=1.40\pm0.04$                                      & XMM=$(4.27^{+0.26}_{-0.24})\times10^{-3}$ & $(5.36^{+0.33}_{-0.30})\times10^{42}$
\enddata
\label{softexcess_merged_MEG_pn_try3_table}
\tablecomments{The normalization is quoted in units of photons cm$^{-1}$ s$^{-1}$ keV$^{-1}$ at~1~keV.}
\end{deluxetable}

\begin{deluxetable}{lllll}
\tablecaption{Fluorescent emission lines detected in the Chandra HETGS spectra}
\tablehead{Line       & Energy (keV)              & $\sigma$ (eV)       & Width (km s$^{-1}$) & Equivalent width (eV)}
\startdata
Fe~K$\alpha$ (narrow) & $6.397\pm0.007$           & $19^{+13}_{-9}$     & $900^{+600}_{-400}$    & $81^{+27}_{-30}$     \\
Fe~K$\alpha$ (broad)  & $6.397\pm0.007$           & $96^{+71}_{-47}$    & $4500^{+3300}_{-2200}$ & $63^{+39}_{-40}$    \\
S K$\alpha$           & $2.303\pm0.008$           & As Si K$\alpha$     & \nodata                & $5.9^{+3.6}_{-3.2}$ \\
Si K$\alpha$          & $1.740\pm0.002$           & $3.6^{+2.4}_{-2.0}$ & $600^{+400}_{-300}$    & $5.6^{+2.0}_{-1.8}$ \\
Ca K$\alpha$          & $3.691^{+0.007}_{-0.006}$ & As Si K$\alpha$     & \nodata                & $4.6^{+3.5}_{-3.0}$ \\
Ar K$\alpha$          & $2.951\pm0.003$           & As Si K$\alpha$     & \nodata                & $6.1^{+4.4}_{-4.3}$ 

\enddata
\label{kalphalines}
\tablecomments{Errors on the energy and width of the lines are quoted as 90\% confidence for two interesting parameters (i.e. $\Delta\chi^2$=2.3). The errors on the equivalent width take into account the 90\% confidence for one interesting parameter associated with the normalization of the Gaussian line only.}
\end{deluxetable}

\begin{deluxetable}{ll}
\tablecaption{Best-fitting Parameters of Multiply Absorbed Power Law Model}
\tablehead{Component & Value}
\startdata
$N_{\rm H, 1}$ & $(1.28\pm 0.13)\times10^{23}$  \\ 
               & $c_1=0.32\pm 0.01$               \\ \hline
$N_{\rm H, 2}$ & $(2.76\pm 0.03)\times10^{22}$  \\ 
               & $c_2=0.96\pm 0.001$               \\ \hline
$N_{\rm H, 3}$ & $(7.68\pm 1.04)\times10^{20}$  \\ 
               & $c_3=1$               \\ \hline
PL             & $\Gamma=1.74\pm 0.05$          \\ 
               & norm$_{MEG}$=$(1.23\pm0.02)\times10^{-2}$ \\  
               & norm$_{HEG}$=$(1.18\pm0.02)\times10^{-2}$ \\ 
               & norm$_{XMM}$=$(0.87\pm0.02)\times10^{-2}$ 
\enddata
\label{multipleabsorber_tab}
\tablecomments{Errors quoted 90\% confidence for one interesting parameter. Note that there are significant degeneracies between these parameters.}
\end{deluxetable}

\end{document}